\documentclass[sigconf,natbib=true,anonymous=false]{acmart}

\AtBeginDocument{%
  \providecommand\BibTeX{{%
    \normalfont B\kern-0.5em{\scshape i\kern-0.25em b}\kern-0.8em\TeX}}}

\copyrightyear{2023}
\acmYear{2023}
\setcopyright{acmlicensed}\acmConference[SIGIR '23]{Proceedings of the 46th International ACM SIGIR Conference on Research and Development in Information Retrieval}{July 23--27, 2023}{Taipei, Taiwan}
\acmBooktitle{Proceedings of the 46th International ACM SIGIR Conference on Research and Development in Information Retrieval (SIGIR '23), July 23--27, 2023, Taipei, Taiwan}
\acmPrice{15.00}
\acmDOI{10.1145/3539618.3591876}
\acmISBN{978-1-4503-9408-6/23/07}

\PassOptionsToPackage{usenames,dvipsnames}{xcolor}
\usepackage{bm}
\usepackage{enumitem}
\usepackage{tikz}
\usepackage{tikz-qtree}
\usepackage{pgfplots}
\usepackage{subfigure}
\usepackage{adjustbox}
\usepackage{multirow}
\usepackage[normalem]{ulem}
\usepackage{soul}
\usepackage{url}
\usepackage[ruled,vlined,linesnumbered]{algorithm2e}
\usepackage{xcolor}

\begin{document}

\title{Towards Building Voice-based Conversational  Recommender Systems: Datasets, Potential Solutions, and Prospects}
\renewcommand{\shorttitle}{Towards Building Voice-based Conversational Recommender Systems}


\author{Xinghua Qu}
\affiliation{
\institution{Bytedance AI Lab}
\city{}
\country{Singapore}
}

\author{Hongyang Liu}\authornote{Co-first authors}
\affiliation{
\institution{Bytedance AI Lab}
\city{Shanghai}
\country{China}
}

\author{Zhu Sun}
\authornote{Corresponding author: sunzhuntu@gmail.com}
\affiliation{%
  \institution{Institute of High Performance Computing, Centre for Frontier AI Research, A*STAR}
  \city{}
  \country{Singapore}
}

\author{Xiang Yin}
\affiliation{%
  \institution{Bytedance AI Lab}
  \city{Shanghai}
  \country{China}
}

\author{Yew Soon Ong}
\affiliation{%
  \institution{A*STAR Centre for Frontier AI Research and  Nanyang Technological University}
  \city{}
  \country{Singapore}
}

\author{Lu Lu}
\affiliation{%
  \institution{Bytedance AI Lab}
  \city{Hangzhou}
  \country{China}
}

\author{Zejun Ma}
\affiliation{%
  \institution{Bytedance AI Lab}
  \city{Beijing}
  \country{China}
}

\renewcommand{\shortauthors}{Xinghua Qu et al.}


\begin{abstract}
Conversational recommender systems (CRSs) have become crucial emerging research topics in the field of RSs, thanks to their natural advantages of explicitly acquiring user preferences via interactive conversations and revealing the reasons behind recommendations. However, the majority of current CRSs are text-based, which is less user-friendly and may pose challenges for certain users, such as those with visual impairments or limited writing and reading abilities. Therefore, \textit{for the first time}, this paper investigates the potential of voice-based CRS (VCRSs) to revolutionize the way users interact with RSs in a natural, intuitive, convenient, and accessible fashion. To support such studies, we create two VCRSs benchmark datasets in the e-commerce and movie domains, after realizing the lack of such datasets through an exhaustive literature review. Specifically, we first empirically verify the benefits and necessity of creating such datasets. 
Thereafter, we convert the user-item interactions to text-based conversations through the ChatGPT-driven prompts for generating diverse and natural templates, and then synthesize the corresponding audios via the text-to-speech model. Meanwhile, a number of strategies are delicately designed to ensure the naturalness and high quality of voice conversations. On this basis, we further explore the potential solutions and point out possible directions to build end-to-end VCRSs by seamlessly extracting and integrating voice-based inputs, thus delivering performance-enhanced, self-explainable, and user-friendly VCRSs. Our study aims to establish the foundation and motivate further pioneering research in the emerging field of VCRSs. This aligns with the principles of explainable AI and AI for social good, viz., utilizing technology's potential to create a fair, sustainable, and just world. Our codes and datasets are available on GitHub (\url{https://github.com/hyllll/VCRS}). 
\end{abstract}

\ccsdesc[500]{Information systems~Recommender systems}

\keywords{Voice-based Recommendation; Conversational Recommender Systems; Explainable AI; AI for Social Good}


\maketitle

\section{Introduction}
\begin{figure*}
    \centering
    \subfigure[Conversation in HOOPS~\cite{fu2021hoops}]{
    \includegraphics[width=0.257\textwidth]{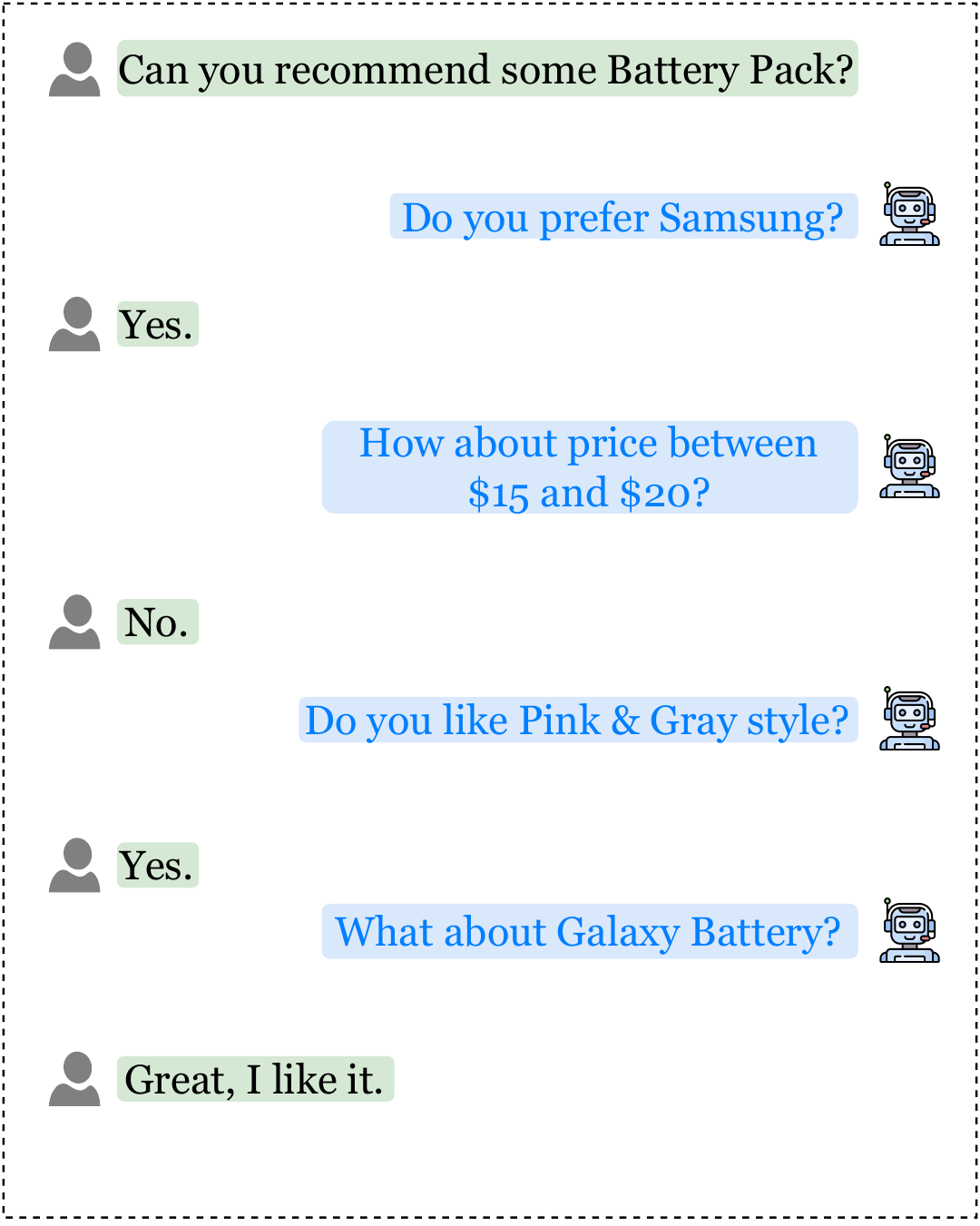}
    }\hspace{0.2in}
    \subfigure[Conversation in Coat]{
    \includegraphics[width=0.27\textwidth]{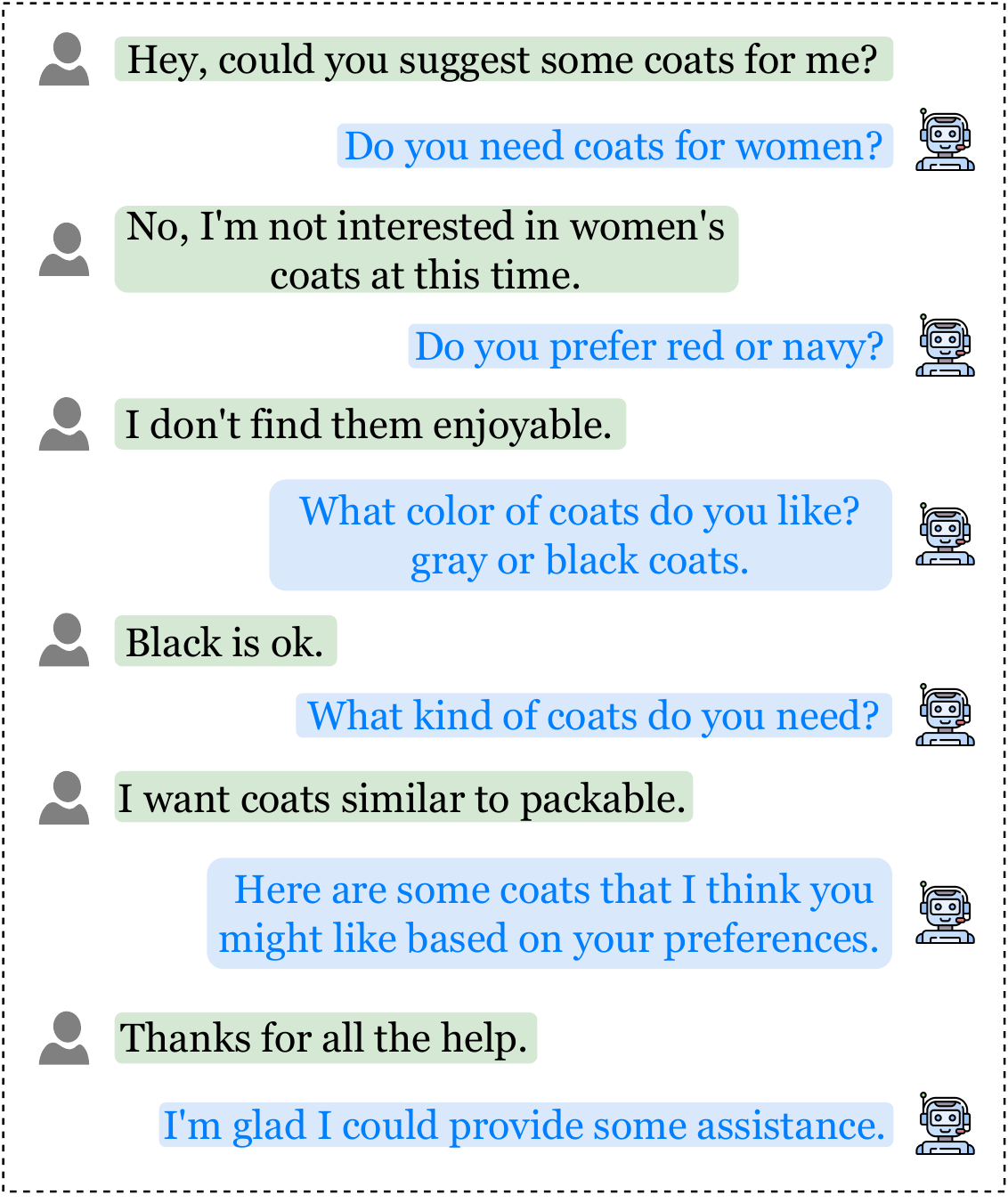}
    }\hspace{0.2in}
    \subfigure[Conversation in OpenDialKG~\cite{moon2019opendialkg}]{
    \includegraphics[width=0.263\textwidth]{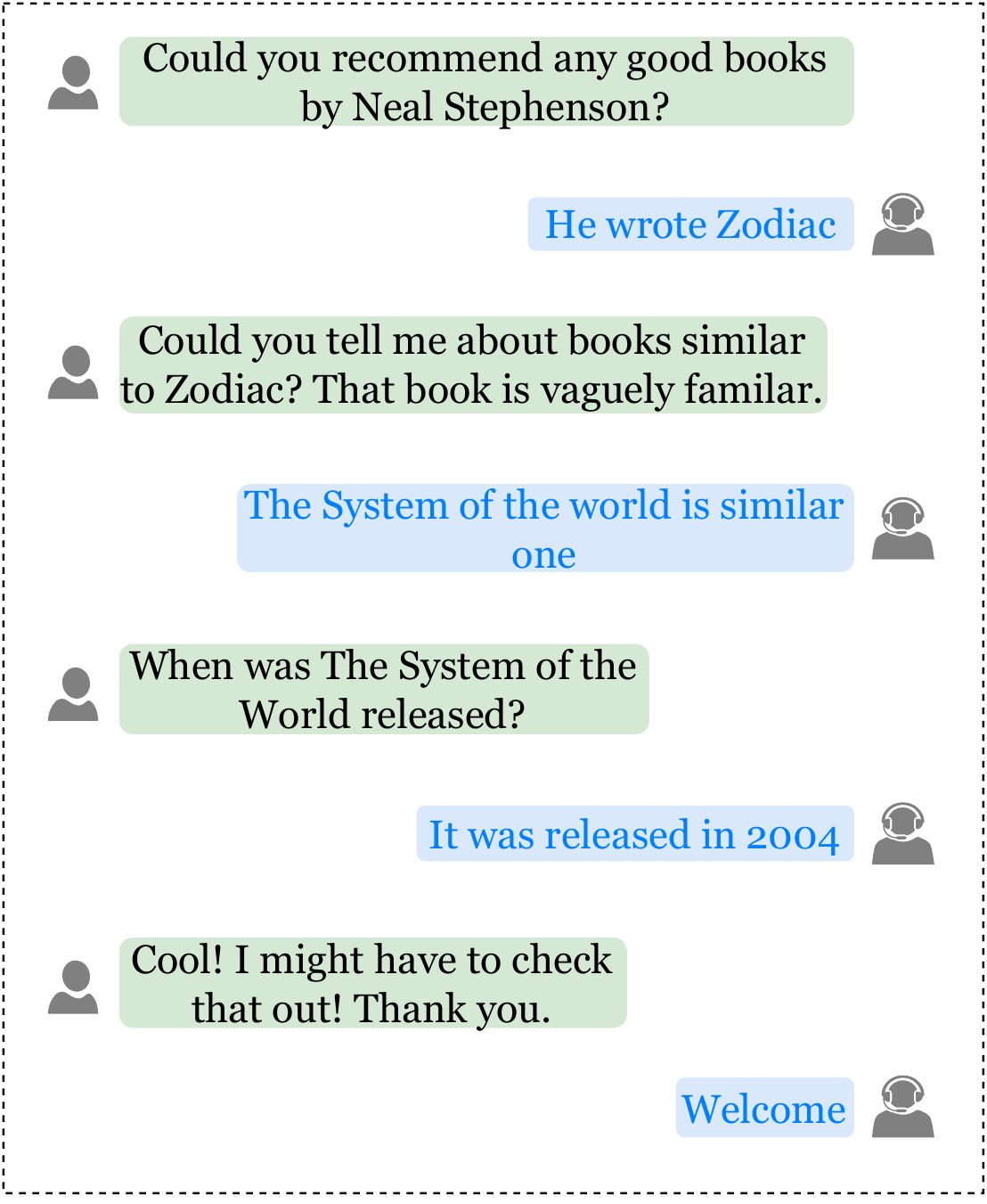}
    }
    \vspace{-0.15in}
    \caption{Running examples of text-based conversations in different scenarios: (a) simulated dialogue in HOOPS; (b) simulated dialogue by us with the Coat dataset; and (c) generated in OpenDialKG via real human dialogue.}
    \label{fig:conversation-example}
    \vspace{-0.1in}
\end{figure*}

Conversational recommender systems (CRSs) have recently gained significant attention in the field of recommender systems (RSs), as evidenced by a number of studies~\cite{gao2021advances,jannach2021survey}. The natural advantages of CRSs, such as the ability to acquire user preferences through interactive conversations and provide intuitive interpretation for recommendations, align with the growing trend of explainable AI. However, the majority of research on CRSs to date~\cite{zhang2022conversation,deng2021unified,li2018towards} mainly focuses on text-based conversations between users and agents, also known as TCRSs, as illustrated in Figure~\ref{fig:conversation-example}. Nonetheless, TCRSs may not be as user-friendly or engaging for certain individuals, such as those with vision impairments or limited writing and reading abilities. The World Health Organization's 2022 statistics\footnote{\url{www.who.int/news-room/fact-sheets/detail/blindness-and-visual-impairment}} indicate that at least 2.2 billion people worldwide have a vision impairment. Moreover, there are many languages that lack a corresponding writing system \cite{nettle2000vanishing}, where the text-based conversation becomes infeasible. As such, there is a pressing need to explore alternative modalities, such as voice-based CRSs (VCRSs)\footnote{or speech-based CRSs.}, to make the benefits of CRSs more accessible to all users.

Compared with text, verbal conversations 
with voice are more direct, natural, convenient, and time-saving. 
As disclosed by a recent report\footnote{\url{www.asmag.com/showpost/26648.aspx}}, voice-based shopping is expected to jump to 40 billion US dollars. 
Meanwhile, another report\footnote{\url{www.hellorep.ai/blog/what-is-voice-commerce}} shows that over 8 billion voice assistants are going to be in use worldwide by 2023, where more than half of these users use them for online shopping. 
Without exception, the utilization of voice inputs in CRSs (i.e., VCRSs) can offer a range of benefits. \textit{Firstly}, the user experience is more direct, engaging, and efficient, as users are less likely to be distracted or lost browsing through products. With VCRSs, users can ask exactly what they want and make orders conveniently\footnotemark[4]. 
\textit{Secondly}, VCRSs are more conversational in nature, and allow for the inclusion of both objective deep features and subjective features in user's queries~\cite{kang2017understanding}. \textit{Thirdly}, voice inputs can convey more information than text inputs, such as user age, gender, and accent, which enables VCRSs to deliver more accurate and transparent recommendations. \textit{Last but not least}, VCRSs are more user-friendly and accessible, particularly for individuals with difficulties in writing and reading.

Given the advantages aforementioned, our long-term goal is to develop end-to-end VCRSs that can revolutionize the way users interact with RSs in a natural, intuitive, convenient, and accessible fashion. This aligns with the principles of explainable AI and AI for social good. With this goal in mind, however, we find that very limited existing studies focused on this topic. To the best of our knowledge, only one work from Amazon~\cite{mairesse2021learning} is working on VCRSs, but it merely focuses on dialogue management for determining the best question to ask or the best item to recommend throughout the conversion, thus does not touch how to properly handle and integrate voice inputs into CRSs. \textit{More importantly}, there is even no public dataset available for conducting research on VCRSs. Therefore, our primary target is to fulfill this gap by providing VCRSs benchmark datasets that future work can mount on. 

In doing so, we first empirically verify the benefits and necessity of voice-based conversations in the context of RSs, which explicitly supports our motivation for VCRSs. 
On this basis, we then construct two benchmark datasets for VCRSs in the domains of e-commerce and movies. Specifically, the constructed VCRSs benchmark datasets are built upon two public datasets, i.e., Coat\footnote{\url{www.cs.cornell.edu/~schnabts/mnar/}} and MovieLens-1M (ML-1M)\footnote{\url{grouplens.org/datasets/movielens/1m/}}, where the user-item interaction records are enriched and converted to text-based conversations first through ChatGPT-driven prompts. Thereafter, the generated text conversations are further synthesized as speech using a well-trained conditional variational autoencoder (a.k.a. VITS~\cite{kim2021conditional}). Based on the constructed datasets, we further explore potential solutions to extract and exploit features from such voice inputs to improve the performance of SOTA recommendation algorithms. Last but not least, possible directions are also pointed out to build end-to-end, performance-enhanced, self-explainable, and user-friendly VCRSs. 

Our main contributions lie four-fold. \textbf{(1)} We empirically verify the advantages and necessity of involving the modality of voice in CRSs. \textbf{(2)} Based on ChatGPT-driven prompts and neural speech synthesis, we create two VCRS benchmark datasets that future work in this area can follow. \textbf{(3)} We explore potential solutions based on the generated VCRSs benchmark datasets, which show significant performance improvement via involving voice-related side information. \textbf{(4)} We point out the possible directions for building end-to-end, performance-enhanced, self-explainable, and user-friendly VCRSs. 

\section{Related Work}
The conversational recommender systems (CRSs)~\cite{chen2019towards, xu2021adapting} has been becoming a hot topic in both industry and academy, due to the two significant advantages brought by conversations, viz., 1) enabling to dynamically capture the user's immediate preferences~\cite{zhang2022conversation}, and 2) revealing the reasons behind the recommendation for explainability.
In this section, we first provide an overview of the related works from the angle of the data source that current CRSs are built upon, mainly categorized by real human dialogue and simulated dialogue. After that, we also review studies related to voice-based CRSs.  

\smallskip\noindent\textbf
{Real Human Dialogue for CRSs}.
Plenty of studies~\cite{iovine2019dataset,narducci2018improving,xu2020user} on CRSs are built on real human dialogues. In general, they initially create specific dialogue tasks and then recruit actual users to engage in natural language conversations. 
Typically,
these conversations are carried out by hiring users on crowd-sourcing platforms like Amazon Mechanical Turk (AMT; \url{www.mturk.com/}).
For instance, in 
ReDial~\cite{li2018towards}, a paired mechanism is used where one person acts as a recommendation seeker and the other as the agent. Such collected datasets are thereafter 
used to train a neural network for conversational movie recommendations. There are also similar studies, 
e.g., 
INSPIRED~\cite{hayati2020inspired}, GoRecDial~\cite{kang2019recommendation}, OpenDialKG~\cite{moon2019opendialkg} and DuRecDial~\cite{liu2020towards}, following the same crowd-sourcing pipeline for data collection.
Besides, some 
recent CRS approaches are evaluated on the above human dialogue datasets, for example, UPCR~\cite{ren2022variational}, TSCR~\cite{zou2022improving}, UniCRS~\cite{wang2022towards}, 
UCCR~\cite{li2022user}, DICR~\cite{zhou2022aligning} and 
C$^2$-CRS~\cite{zhou2022c2}.

Although these real human dialogues collected can help in training CRSs, they are usually
time-consuming, laborious, and expensive. For instance, OpenAI even hires their data collection contributors as full-time employees. 
Consequently, the sizes of such datasets are limited due to the high cost. Thus, an alternative and more economical way of CRSs data collection is by simulation, especially when a larger-scale dataset is required.

\smallskip\noindent\textbf
{Simulated Dialogue for CRSs}.
Another line of research on CRSs simulates dialogue for each user-item interaction record with the aid of side information, e.g., item features and knowledge graphs~\cite{ren2021learning,zhang2022conversation}. In particular, the conversation simulation is formulated as a slot-filling problem, where the asked features are filled in manually pre-defined templates. In this sense, they serve as upper-bound studies of real applications as the errors in language understanding and generation are not included.
%
For instance, CRM~\cite{sun2018conversational}
delexicalizes user utterances collected from AMT into templates~\cite{wen2016network}, and determines the asked features via a deep policy network.
Similar works also include EAR~\cite{lei2020estimation}, CART~\cite{zhang2020conversational}, and 
FPAN~\cite{xu2021adapting}.
Following the same paradigm, 
SCPR~\cite{lei2020interactive}, HOOPS~\cite{fu2021hoops}, KBQG~\cite{ren2021learning},
GPR~\cite{zhao2022two}
and HICR~\cite{tu2022conversational}
further leverage (knowledge) graphs to guide the generation of the asked features or recommended items.
Recently, to further foster the scalability and generality of CRSs, 
UNICORN~\cite{deng2021unified} and   CRIF~\cite{hu2022learning} replace the separated conversation and recommendation components in previous studies with a unified policy learning process. 
On this basis, 
MCMIPL~\cite{zhang2022multiple} takes a step further to enable the agent to generate multiple-choice questions, rather than binary yes/no questions, about specific features. 
%
%

However, all these works only focus on simulated text-based CRSs, thus being different from us for voice-based ones. Moreover, in this paper, we involve ChatGPT (a more advanced large language model) to generate 
conversation templates and enable multiple-choice questions, so as to ensure the naturalness, multiplicity, and rationality of the simulated conversations.

\begin{table}[t]
\footnotesize
  \caption{Statistics of datasets (InterA. and ConverS. are short names for Interactions and Conversations, respectively). 
  }
  \label{tab:statistics}
  \vspace{-0.15in}
  \begin{tabular}{l|cc|cc}
    \toprule
    &\multicolumn{2}{c|}{\textbf{Coat}}  &\multicolumn{2}{c}{\textbf{ML-1M}} \\\cline{2-5}
    &text-based &voice-based &text-based & voice-based\\\midrule
    \#Users & 290 & 290 &5,818  &5,818\\
    \#Items & 300 & 300 &3,250 &3,250\\
    \#InterA./ConverS. &6,960 &6,960 &908,308 &10,000 \\\hline
    User Features  & \multicolumn{2}{c|}{gender, age, location} & \multicolumn{2}{c}{gender, age}  \\
    Item Features & \multicolumn{2}{c|}{gender, jacket type, color}  & \multicolumn{2}{c}{genre, actor, director, country} \\
  \bottomrule
\end{tabular}
\vspace{-0.15in}
\end{table}

\smallskip\noindent\textbf
{Voice-based CRSs}.
According to the literature review, only one study~\cite{mairesse2021learning} has considered VCRSs, in comparison to the wealth of related work on text-based CRSs. Nonetheless, it merely focuses on policy learning to determine the best questions to ask, which has been applied to Amazon Music yet without providing an open-source dataset and models. As a result, there is currently no available public dataset for research on VCRSs.
Besides, another work~\cite{zhang2020web}, though involving voice-based recommendation, aims to transfer customers' shopping patterns from the web to the voice platform, thus being different from our goal of CRSs directly on voices.

As previously mentioned, VCRSs offer a more friendly interaction with users and are more explainable for recommendations provided. VCRSs allow users to engage in a more natural and intuitive way, and users can also ask follow-up questions, leading to a more dynamic and personalized experience. Additionally, the use of voice-based communication can help to overcome barriers in text-based communication for individuals with certain disabilities or those who are not literate. Therefore, there is great potential for VCRSs to improve the user experience in various domains, and further research is needed to develop effective approaches for VCRSs.

\section{The Constructed Datasets}
In this section, we first verify the advantages and necessity of constructing voice-based conversation datasets in the domain of VCRSs. Then, we introduce the core steps involved in the data construction process, so as to deliver two benchmark datasets that future work in this area can follow.

\subsection{Advantages and Necessity}\label{subsec:advantages-and-necessity}
Besides the user-friendly and convenient fashion provided, verbal conversations with voices also convey more information (such as age, gender, accent, and emotional status) than pure text-based conversations. Moreover, we believe that extracting those side information from voices is necessary and more reliable due to two reasons: 1) some side information can only come from voices, such as accent and emotion status, viz., the user registration behind the backend system is hard to obtain such detailed and dynamic information; 2) even though some side information (e.g., gender and age) can be obtained, they usually have data missing and fake information issues, namely user registration information is typed in randomly or restricted to access. In contrast, the human voice is honest in delivering such information, especially when a voice-based conversation is going on. We hypothesize that such side information is essential for more accurate recommendations.  

\begin{figure*}[t]
    \centering \includegraphics[width=0.95\textwidth]{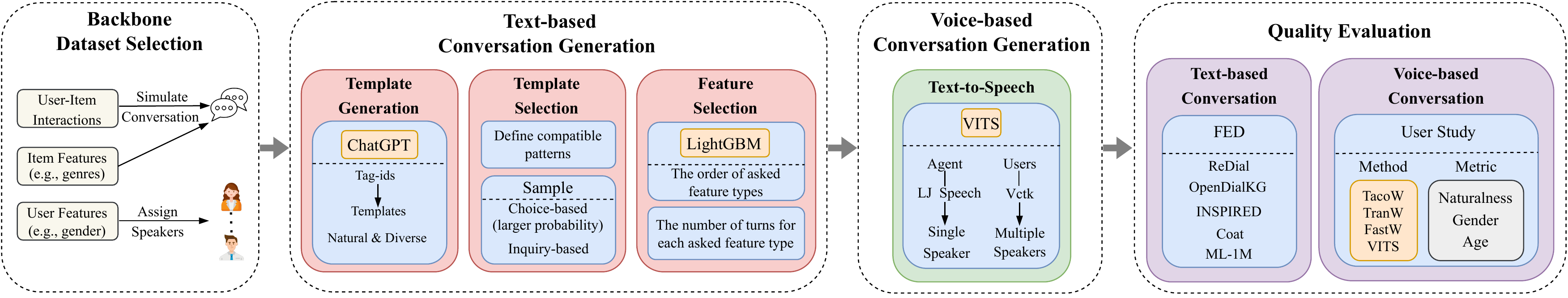}
    \vspace{-0.1in}
    \caption{The core steps to construct voice-based conversation in the context of RSs.}
    \label{fig:core-steps}
    \vspace{-0.15in}
\end{figure*}


To verify such a hypothesis, we first analyze the recommendation performance improvement by involving the side information that can be extracted from voice inputs. 
Given that there is no existing VCRS algorithm and dataset available for the hypothesis test, we alternatively achieve this goal by the analysis on public datasets using feature-based recommendation algorithms. 
We figure out two public datasets (i.e., Coat and ML-1M) with extra user features (such as gender, age, beside user-item interactions) provided among many without. These extra user features can be regarded as the side information correctly extracted from voice conversations\footnote{We empirically verified that by using SOTA classification models, these features can be extracted from voice inputs with an accuracy of around 90\%.}.
More specifically, Coat contains user-clothing interaction records, user features (i.e., gender, age, and location), and clothing features (i.e., gender, jacket type, and color).
ML-1M comprises records of user-movie interactions, user features (i.e., age and gender), and movie features (i.e., genre, actor, director, and country) crawled from IMDB\footnote{\url{https://www.imdb.com/}}. 
For ML-1M, we filter out movies with missing features on IMDB and features involving non-English words (e.g., director names). The statistics of the two datasets after pre-processing are summarized in Table~\ref{tab:statistics} (refer to the `text-based' columns). 

\begin{table*}[t]
\centering
\footnotesize
\caption{Performance of SOTA feature-based recommenders on Coat and ML-1M, where each experiment is repeatedly executed five times, and the mean $\pm$ standard deviation values (\%) are reported; the best results are highlighted in bold, and the worst results are marked with `*'; and the row `\textit{drop}' indicates the relative decrease comparing the best and worst results.}
\vspace{-0.15in}
\addtolength{\tabcolsep}{4pt}
\begin{tabular}{ll|ccc|ccc}
\toprule
                                            &                  & \multicolumn{3}{c|}{FM}                                                           & \multicolumn{3}{c}{WD}                                                           \\ \cline{3-8} 
                                            &                  & Precision@10              & Recall@10                 & NDCG@10                   & Precision@10             & Recall@10                 & NDCG@10                   \\ \hline
\multicolumn{1}{c|}{\multirow{6}{*}{Coat}}  & all features     & \textbf{4.6968} $\pm$ 0.2168  & \textbf{10.1197} $\pm$ 1.0912 & \textbf{20.1047} $\pm$ 0.7627 & \textbf{4.9059} $\pm$ 0.3346 & \textbf{10.1599} $\pm$ 0.9260 & \textbf{19.3744} $\pm$ 1.6603 \\
\multicolumn{1}{c|}{}                       & w/o gender       & 4.2577 $\pm$ 0.4311           & 9.1751 $\pm$ 0.6185           & 18.1176 $\pm$ 2.0496          & 4.8919 $\pm$ 0.3133          & 9.9916 $\pm$ 0.9830           & 18.5858 $\pm$ 1.5083          \\
\multicolumn{1}{c|}{}                       & w/o age          & 4.6411 $\pm$ 0.3514           & 9.4634 $\pm$ 1.1302           & 19.1396 $\pm$ 1.5554          & 4.8362 $\pm$ 0.3828          & 9.8784 $\pm$ 0.8115           & 17.3254 $\pm$ 1.9464          \\
\multicolumn{1}{c|}{}                       & w/o location     & 4.5505 $\pm$ 0.4749           & 9.4845 $\pm$ 0.9145           & 18.7961 $\pm$ 1.1137          & 4.8780 $\pm$ 0.3107          & 10.2198 $\pm$ 0.4233          & 19.0911 $\pm$ 1.4939          \\
\multicolumn{1}{c|}{}                       & w/o all features & 4.0766* $\pm$ 0.3703          & 8.5447* $\pm$ 0.6432          & 16.8249* $\pm$ 2.0686         & 4.5573* $\pm$ 0.6276         & 9.1530* $\pm$ 1.4369          & 17.2747* $\pm$ 2.0054         \\
\multicolumn{1}{c|}{}                       & \textit{drop}    & 13.2047\%                   & 15.5637\%                     & 16.3136\%                   & 7.1057\%                   & 9.9105\%                    & 10.8375\%                   \\ \hline
\multicolumn{1}{l|}{\multirow{5}{*}{ML-1M}} & all features     & \textbf{45.8643} $\pm$ 0.4120 & \textbf{8.6708} $\pm$ 0.2912  & \textbf{66.1315} $\pm$ 0.6313 & \textbf{9.8166} $\pm$ 0.3319 & \textbf{1.0836} $\pm$ 0.1974  & \textbf{24.6374} $\pm$ 1.0423 \\
\multicolumn{1}{l|}{}                       & w/o gender       & 45.7188 $\pm$ 0.2070          & 8.5023 $\pm$ 0.2278           & 65.5473 $\pm$ 0.5163          & 9.4388 $\pm$ 0.2976          & 0.8407 $\pm$0.0599            & 23.5227 $\pm$ 0.8462          \\
\multicolumn{1}{l|}{}                       & w/o age          & 45.5743 $\pm$ 0.2310          & 8.5392 $\pm$ 0.1998           & 65.5267 $\pm$ 0.5349          & 9.4623 $\pm$ 0.7633          & 0.9311 $\pm$ 0.1542           & 23.6480 $\pm$ 1.7351          \\
\multicolumn{1}{l|}{}                       & w/o all features & 45.5037* $\pm$ 0.1903         & 8.3053* $\pm$ 0.0997          & 65.3110* $\pm$ 0.3989         & 9.1353* $\pm$ 0.7609         & 0.9129* $\pm$ 0.1965          & 21.1502* $\pm$ 2.2782         \\
\multicolumn{1}{l|}{}                       & \textit{drop}    & 0.7862\%                    & 4.2153\%                    & 1.2407\%                    & 6.9403\%                   & 15.7530\%                   & 14.1541\%\\   
\bottomrule
\end{tabular}
\label{tab:ablation-study}
\vspace{-0.15in}
\end{table*}

Mounted on Coat and ML-1M, two SOTA feature-based recommendation algorithms, i.e., factorization machine (FM)~\cite{rendle2010factorization} and wide\&deep (WD)~\cite{cheng2016wide} are considered. Subsequently, we conduct an extensive ablation study to verify the benefits of extra user features for more accurate recommendations\footnote{All detailed experimental settings in this paper are available on our GitHub.}.   
Table~\ref{tab:ablation-study} presents the performance of FM and WD on Coat and ML-1M by evaluating their generated top-$K$ recommendation list ($K=10$). Several major findings can be noted: (1) user features indeed help improve the recommendation accuracy, for instance, there is a 
16.3\% performance drop by removing all extra user features from FM on Coat regarding NDCG@10;
(2) the impacts vary a lot across features, models and domains, for example, user gender affects the performance of FM on Coat most; whilst age is the most impactful feature for WD on Coat. 
and (3) overall, the huge relative performance drops (i.e., 7.00\%, 11.36\%, and 10.64\% w.r.t. Precision, Recall, and NDCG across the two domains on average) firmly indicate that significant enhancements in performance would be achieved if we could extract these extra user features from the voice inputs, confirming the benefits and necessity of creating such voice-based datasets for VCRSs.     

\subsection{Data Construction} 
Bearing the advantages and necessity of voice-based datasets for VCRSs in mind, this paper aims to create such datasets for future work in VCRSs,
given there is no existing one. To ensure better data quality, we first construct text-based conversions and then adopt SOTA text-to-speech (TTS) models to convert text into audio by assigning appropriate speakers based on user profiles. 
Keeping this in view, we propose a dataset creation pipeline based on ChatGPT-driven templates and neural speech synthesis as depicted in Figure~\ref{fig:core-steps}. In specific, our VCRSs benchmark dataset creation task includes four steps (1) backbone dataset selection; (2) text-based conversation generation; (3) voice-based conversation generation; and (4) quality evaluation. The detail of each step is illustrated as follows. 

\definecolor{airforceblue}{rgb}{0.36, 0.54, 0.66}
\definecolor{aliceblue}{rgb}{0.94, 0.97, 1.0}
\definecolor{alizarin}{rgb}{0.82, 0.1, 0.26}
\definecolor{almond}{rgb}{0.94, 0.87, 0.8}
\definecolor{amber}{rgb}{1.0, 0.75, 0.0}
\definecolor{amber(sae/ece)}{rgb}{1.0, 0.49, 0.0}
\definecolor{amethyst}{rgb}{0.6, 0.4, 0.8}
\definecolor{antiquebrass}{rgb}{0.8, 0.58, 0.46}
\definecolor{antiquefuchsia}{rgb}{0.57, 0.36, 0.51}
\definecolor{applegreen}{rgb}{0.55, 0.71, 0.0}
\definecolor{apricot}{rgb}{0.98, 0.81, 0.69}
\definecolor{arylideyellow}{rgb}{0.91, 0.84, 0.42}
\definecolor{ashgrey}{rgb}{0.7, 0.75, 0.71}
\definecolor{atomictangerine}{rgb}{1.0, 0.6, 0.4}
\definecolor{aureolin}{rgb}{0.99, 0.93, 0.0}
\definecolor{azure(colorwheel)}{rgb}{0.0, 0.5, 1.0}
\definecolor{babypink}{rgb}{0.96, 0.76, 0.76}
\definecolor{bluebell}{rgb}{0.64, 0.64, 0.82}
\definecolor{brightlavender}{rgb}{0.75, 0.58, 0.89}
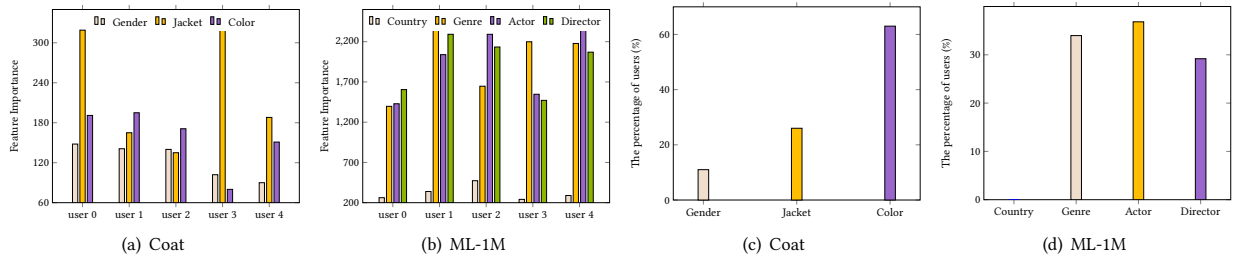
\begin{figure*}[t]
\centering
\subfigure[Coat]{
\begin{tikzpicture}[scale=0.4]
\pgfplotsset{%
    width=0.55\textwidth,
    height=0.45\textwidth
}
\begin{axis}[
    ybar,
    bar width=5pt,
    ylabel={Feature Importance},
    ylabel style ={font = \Large},
    xlabel style ={font = \LARGE},
    enlarge x limits={abs=1.0cm},
    scaled ticks=false,
    tick label style={/pgf/number format/fixed, font=\Large},
    ymin=60, ymax=350,
    symbolic x coords={user 0, user 1, user 2, user 3, user 4},
    xtick=data,
    ytick={60,120,180,240,300},
    legend style={at={(0.5,0.98)}, anchor=north,legend columns=3, column sep=0.2cm, draw=none, font=\Large},
]
\addplot [fill=almond]coordinates {
(user 0, 148) (user 1, 141)
(user 2, 140) (user 3, 102)
(user 4, 90)};
\addplot [fill=amber]coordinates {
(user 0, 319) (user 1, 165)
(user 2, 135) (user 3, 334)
(user 4, 188)}; 
\addplot [fill=amethyst]coordinates {
(user 0, 191) (user 1, 195)
(user 2, 171) (user 3, 80)
(user 4, 151)};
\legend{Gender, Jacket, Color}
\end{axis}
\end{tikzpicture}
}
\subfigure[ML-1M]{
\begin{tikzpicture}[scale=0.4]
\pgfplotsset{%
    width=0.55\textwidth,
    height=0.45\textwidth
}
\begin{axis}[
    ybar,
    bar width=5pt,
    ylabel={Feature Importance},
    ylabel style ={font = \Large},
    xlabel style ={font = \LARGE},
    enlarge x limits={abs=1.0cm},
    scaled ticks=false,
    tick label style={/pgf/number format/fixed, font=\Large},
    ymin=200, ymax=2600,
    symbolic x coords={user 0, user 1, user 2, user 3, user 4},
    xtick=data,
    ytick={200,700,1200,1700,2200},
    legend style={at={(0.5,0.98)}, anchor=north,legend columns=4, column sep=0.2cm, draw=none, font=\Large},
]
\addplot [fill=almond] coordinates {
(user 0, 262) (user 1, 339)
(user 2, 473) (user 3, 241)
(user 4, 289)};
\addplot [fill=amber] coordinates {
(user 0, 1397) (user 1, 2490)
(user 2, 1647) (user 3, 2199)
(user 4, 2178)}; 
\addplot [fill=amethyst] coordinates {
(user 0, 1429) (user 1, 2039)
(user 2, 2292) (user 3, 1547)
(user 4, 2390)};
\addplot [fill=applegreen] coordinates {
(user 0, 1605) (user 1, 2292)
(user 2, 2134) (user 3, 1471)
(user 4, 2070)};
\legend{Country, Genre, Actor, Director}
\end{axis}
\end{tikzpicture}
}
\subfigure[Coat]{
\begin{tikzpicture}[scale=0.4]
\pgfplotsset{%
    width=0.55\textwidth,
    height=0.45\textwidth
}
\begin{axis}[
    ybar,
    bar width=10pt,
    ylabel={The percentage of users (\%)},
    ylabel style ={font = \Large},
    xlabel style ={font = \Large},
    enlarge x limits={abs=1.0cm},
    scaled ticks=false,
    tick label style={/pgf/number format/fixed, font=\Large},
    ymin=0, ymax=70,
    xtick={1,2,3},
    xticklabels={Gender, Jacket, Color},
    ytick={0,20,40,60},
    every axis plot/.append style={
          bar shift=0pt,
        }
]
\addplot [fill=almond] coordinates {(1, 11)}; 
\addplot [fill=amber] coordinates {(2, 26)}; 
\addplot [fill=amethyst] coordinates {(3, 63)}; 
\end{axis}
\end{tikzpicture}
}
\subfigure[ML-1M]{
\begin{tikzpicture}[scale=0.4]
\pgfplotsset{%
    width=0.55\textwidth,
    height=0.45\textwidth
}
\begin{axis}[
    ybar,
    bar width=10pt,
    ylabel={The percentage of users (\%)},
    ylabel style ={font = \Large},
    xlabel style ={font = \Large},
    enlarge x limits={abs=1.0cm},
    scaled ticks=false,
    tick label style={/pgf/number format/fixed, font=\Large},
    ymin=0, ymax=40,
    xtick={1,2,3,4},
    xticklabels={Country, Genre, Actor, Director},
    ytick={0,10,20,30},
    every axis plot/.append style={
          bar shift=0pt,
        }
]
\addplot coordinates {(1, 0)};
\addplot [fill=almond] coordinates {(2, 33.98)}; 
\addplot [fill=amber] coordinates {(3, 36.83)}; 
\addplot [fill=amethyst] coordinates {(4, 29.19)}; 
\end{axis}
\end{tikzpicture}
}
\vspace{-0.2in}
\caption{
(a-b) the importance of different feature types w.r.t. five randomly sampled users; and (c-d) the user distribution w.r.t. the feature type of the highest importance values.
}\label{fig:feature-importance}
\vspace{-0.15in}
\end{figure*}

\subsubsection{Backbone Dataset Selection}
Based on the literature review, there are some available text-based conversation datasets in RSs, which contain real human dialogues in different domains, e.g., ReDial~\cite{li2018towards} and INSPIRED~\cite{hayati2020inspired} for movie recommendation; and  OpenDialKG~\cite{moon2019opendialkg} for both movie and book recommendation.  
However, these datasets normally do not contain extra user features, e.g., user age and gender, which makes it infeasible for assigning proper speakers according to user features during audio synthesis. 

Typically, a good candidate for the backbone dataset should contain three types of information: user-item interactions, user features, and item features. Using user-item interactions and item features, we can simulate a text-based conversation between users and agents for recommendation. To further synthesize the simulated conversation into audio, we have to access the user features (e.g., gender and age) to assign proper speakers accordingly. Recalling the preliminary analysis in Section~\ref{subsec:advantages-and-necessity}, we choose Coat and ML-1M as our backbone datasets.


\subsubsection{Text-based Conversation Generation}
Given the selected backbone datasets, we now simulate the dialogue between users and agents to obtain text-based conversations. 
Our study follows the simulated dialogue setting~\cite{zhang2022conversation}, wherein the CRS focuses on simulating
such a scenario instead of real dialog. In this setting, the agent interacts with
a user by asking for specific item features via slot filling~\cite{wiseman2018learning} yet with fixed manual templates (e.g., Do you like \underline{track coat}?). Such a process is repeated multiple times to clarify the user's preference until the user obtains recommendations or chooses to quit~\cite{sun2018conversational}.  
However, such fixed manual templates simulated conversations usually look slightly rigid and less natural as shown in Figure~\ref{fig:conversation-example}(a). To create a natural and diverse conversation similar to real dialogue shown in Figure~\ref{fig:conversation-example}(c), we propose to use ChatGPT-driven templates and the corresponding template and feature selection mechanisms introduced below.

In particular, for each user-item interaction record, our goal is to simulate a multi-turn conversation by inquiring about the user's preference towards features associated with the item until the item is successfully recommended or reaching the maximum number of turns (i.e., $Max\_T$). Following~\cite{zhang2022conversation,lei2020interactive}, we set $Max\_T=10$.          
To achieve this, three key modules are involved, including template generation, template selection, and feature selection.  

\smallskip\noindent\textbf{Template Generation}.
To ensure natural and diverse conversations, we first generate diverse templates for inquiry and response. Particularly, for the agent, given each type of item feature (e.g., jacket type), we vary the way of asking questions by either directly inquiring about users' favorite item feature (inquiry-based with tag-id: 00) or providing single or multiple choices (choice-based with tag-id: 01). 
Below shows some examples of the templates w.r.t. asking for jacket type. 
\begin{itemize}[leftmargin=*]
    \item What kind of jacket do you like? (tag-id: 00)
    \item Are you looking for a specific style of jacket? (tag-id: 00)
    \item Do you prefer \underline{track} jacket over other types of jacket? (tag-id: 01)
    \item What type of jacket do you like? \underline{track} or \underline{trench}? (tag-id: 01)
\end{itemize}
According to the questions asked for each feature type by the agent, the user responds based on templates with a wide range of choices. We now demonstrate some examples of the templates corresponding to the question templates above.
\begin{itemize}[leftmargin=*]
    \item I enjoy \underline{track} jacket. (tag-id: 00)
    \item Absolutely, I'm a big fan of \underline{track} jacket. (tag-id: 00)
    \item Yes, I'm really into it. (tag-id: 01)
    \item I love the look and feel of \underline{track} jacket. (tag-id: 01)
\end{itemize}

Additionally, we also generate templates for the agent and users regarding item recommendations, and opening/closing the conversation. Given the space limit, the full list of templates used in our conversation generation can refer to our GitHub. 
It is noteworthy that all these templates under different tag-ids are produced by ChatGPT
(\url{chat.openai.com/chat}),
a large language model (LLM) released by OpenAI on 30 November 2022. The training of ChatGPT largely follows InstrucGPT~\cite{ouyang2022training}, which utilizes supervised training on human expert labels for question-answering conversations and reinforcement learning with an expert-driven reward model. It has been recognized as one of the most powerful LLMs and embedded into Microsoft's new Bing search engine. Given that, and as the demo shown in Figure~\ref{fig:conversation-example}(b), our proposed ChatGPT-driven templates intuitively indicate the naturalness of our generated conversations.


\smallskip\noindent\textbf{Template Selection}.
Given the well-defined templates, we then need to select the proper ones for both the agent and users at each turn of the conversation. Due to the diversity of the templates, it would inevitably lead to dialogues with mismatched questions and answers, for example, 
\begin{itemize}[leftmargin=*]
    \item[] \textit{Agent}: Are you interested in \underline{men's} or \underline{women's} jacket? 
    \item[] \textit{User}: Yes, I like it.
    \item[] \textit{Agent}: Do you like \underline{red}?
    \item[] \textit{User}: I prefer \underline{bomber} jackets.
\end{itemize}
\noindent
To avoid such an awkward situation and increase the coherence of the conversation, the following strategy is designed. In particular,
we first add a tag for each template of the agent and users (see `tag-id' in the above examples), and define compatible matching patterns for these tags. In the above examples, only the templates with the same `tag-id' from the agent and user sides are compatible. 
Note that, in our study, we have a number of rules to help define compatible patterns with these tag-ids. 
Accordingly, in each turn of the conversation, we will randomly sample one template from the agent side and the user side, respectively, guided by the pre-defined compatible patterns.
This, therefore, preserves the rationality of the generated conversations.
%

Besides, we give a larger probability of sampling choice-based templates (tag-id: 01) more than inquiry-based (tag-id: 00) ones for the agent when posing questions. This is motivated by the intuitive fact that users usually don't have clear intent at the beginning when interacting with the CRSs. Namely, CRSs need to elicit user preference via the multi-turn conversations~\cite{zhou2020towards}.

\smallskip\noindent\textbf{Feature Selection}. 
Given the selected templates, we need to determine the best questions to ask (i.e., asked item features), enabling the system to quickly clarify user preference and make a better recommendation ultimately. Firstly, we need to determine \textit{the order of asked feature types} (e.g., jacket type or color), as different types of features may have different impacts on users' final decision-making. For instance, certain users may prioritize the style of the jacket, whereas others may place a greater emphasis on the jacket's color. 
If more impactful feature types are inquired about earlier, the user preference can be inferred more rapidly. 
Consequently, we exploit the decision tree based method LightGBM~\cite{ke2017lightgbm} to calculate the importance of different feature types. Feature types with lower importance values obtained via LightGBM are closer to the root, indicating more impact on users' final decision-making; thus should be asked at an earlier stage. Figures~\ref{fig:feature-importance}(a-b) visualize the importance of feature types w.r.t. five randomly sampled users on Coat and ML-1M; and Figures~\ref{fig:feature-importance}(c-d) display the user distribution w.r.t. the feature type of the highest importance value. 
The obtained decision tree is finally involved in the text-based conversation generation process to determine the order of asked feature types. 

Thereafter, we need to determine \textit{the number of turns for each feature type being asked}. If the selected agent template is inquiry-based, we can directly obtain the ground-truth item feature value of the asked feature type from the user with one turn only. If it is choice-based, the number of turns depends on when the ground-truth feature value can be sampled. 
To ensure the rationality of the conversation, we follow the interaction distribution w.r.t. feature values of the asked feature type when sampling feature values for choice-based templates as shown in Figure~\ref{fig:feature-distribution}.
However, to avoid endless inquiry we set the maximum number of turns for each feature type being asked as 3 (i.e., $Max\_T_f=3$).
The whole process of feature selection is summarized in Algorithm~\ref{alg:feature-selection}.

\begin{table}[t]
\footnotesize
  \caption{Statistics of speakers or users on three datasets.
  }
  \label{tab:speaker-statistics}
  \addtolength{\tabcolsep}{2pt}
  \vspace{-0.15in}
  \begin{tabular}{l|cc|cc|cc}
    \toprule
    &\multicolumn{2}{c|}{Vctk} &\multicolumn{2}{c|}{Coat}  &\multicolumn{2}{c}{ML-1M} \\\cline{2-7}
    &Male &Female &Male &Female &Male &Female \\
    \midrule
    $< 20$ &5 &8 &1 &1 &805 &298\\
    $[20, 30]$ &40 &53 &65 &48 &1,538 &558\\
    $> 30$ &1 &2 &96 &79 &1,844 &775\\
  \bottomrule
\end{tabular}
\vspace{-0.15in}
\end{table}  

\input{plot/feature-distribution.tex}
\begin{algorithm}[t]
\small
\caption{Feature Selection}
\label{alg:feature-selection}
\KwIn {
feature type set $\mathcal{F}_t$,
$Max\_T$, $Max\_T_f$}
Initialization: $Max\_T=10, Max\_T_f=3$\;
\While{$(t<=Max\_T \cap \mathcal{F}_t \neq  \varnothing)$}
{
    \tcp{determine the order of asked feature types}	
    Select $f_t \in \mathcal{F}_t$ with the lowest importance gained by LightGBM\;
    $\mathcal{F}_t \leftarrow \mathcal{F}_t.remove(f_t)$\;
    Sample one template for the agent based on $f_t$\;
    \tcp{determine the number of turns for $f_t$ being asked}
    \If{(agent\_template\_type == inquiry-based)}
    {
        Sample a compatible template for the user\;
        continue\;
    }
    \If{(agent\_template\_type == choice-based)}
    {
        \For{($t_f=1; t_f<=Max\_T_f; t_f++$)}
        {
            \If{$(t+t_f-1) > Max\_T$} 
            {break\;}
            Sample asked feature values for $f_t$ based on Figure~\ref{fig:feature-distribution}\; 
            Sample a compatible template for the user\;
            \If{(feature\_value == ground\_truth\_feature\_value)}
            {
                break\;
            }
        }
        $t \leftarrow t + t_f -1$\;
    }
    
    %
       
}
\end{algorithm}

\subsubsection{Voice-based Conversation Generation}
With the text-based conversation generated, we proceed to generate the voice-based conversation by adopting the SOTA end-to-end text-to-speech (TTS) model VITS\footnote{\url{github.com/jaywalnut310/vits}}. In particular, for our agent, we utilize VITS trained by the LJ Speech dataset\footnote{\url{keithito.com/LJ-Speech-Dataset/}} containing short audio clips of a single speaker; whereas for users, we exploit VITS trained by the Vctk dataset\footnote{\url{www.udialogue.org/download/cstr-vctk-corpus.html}}, which includes speech data uttered by 109 English speakers with various accents. The statistics of the speakers regarding gender and age are summarized in Table~\ref{tab:speaker-statistics}, where we divide their ages into three groups, i.e., $<20, [20, 30], >30$. Meanwhile, we also display the statistics of the users w.r.t. gender and age on Coat and ML-1M. Thereafter, for each text-based conversation, we match its corresponding user with an appropriate speaker according to their gender and age, 
so as to convert it into a voice-based conversation. Due to the huge volume of conversations on ML-1M, saving a full dataset would cause huge storage requirements (approximately 627 Gb), especially for online open-sourcing. Given that, we open-source a mini version on ML-1M with using $1\times10^4$ conversation audio clips. For accessing the full VCRSs dataset based on ML-1M, we provide a python script interface that can be directly involved in dataloader. Finally, the statistics of the two generated VCRSs datasets are presented in Table~\ref{tab:statistics} (refer to `voice-based' columns).


\subsubsection{Quality Evaluation}
We examine the quality of our constructed datasets from the perspectives of both text-based conversation and voice-based conversation. 

\smallskip\noindent\textbf{Evaluation on the Text-based Conversation}.
To automatically and systematically measure the quality of the generated text-based conversation, we adopt the SOTA \underline{f}ine-grained \underline{e}valuation of \underline{d}ialogue (FED) metric~\cite{mehri2020unsupervised}, an automatic evaluation metric using a massively pre-trained model DialoGPT~\cite{zhang2019dialogpt}. There are three advantages of using FED metric: (1) it does not rely on a ground-truth response, (2) it does not require any training data, and (3) it measures fine-grained dialog qualities at both the turn level (e.g., correct, understandable and interesting) and whole dialog level (e.g., coherent, consistent and diverse). For comparison with our constructed two datasets, 
the same evaluation on three real-world human dialogue datasets (i.e., ReDial~\cite{li2018towards}, OpenDialKG~\cite{moon2019opendialkg}, and INSPIRED~\cite{hayati2020inspired}) is utilized as benchmarks. 
Specifically, each time we take one human dialogue dataset as the benchmark (e.g., ReDial) and calculate the performance 
gap between the benchmark and the rest four (e.g., OpenDialKG, INSPIRED, Coat, and ML-1M). Generally, a smaller performance gap with the benchmark indicates better quality of the corresponding dataset. 

\begin{table}[t]
\footnotesize
  \caption{Evaluation on the text-based conversation. 
  }
  \label{tab:text-evaluation}
  \addtolength{\tabcolsep}{0pt}
  \vspace{-0.15in}
  \begin{tabular}{l|cccc|cc}
    \toprule
    &ReDial &OpenDialKG &INSPIRED &Average &Coat &ML-1M
    \\\midrule
    Score &3.1781 & 3.1402 &3.2447 &3.1877 &3.1707 &3.1945\\\hline
    Gap-R&-- &0.0379 &0.0666* &0.0096 &\textbf{0.0074} &0.0164\\
    Gap-O&0.0379 &-- &0.1045* &0.0475 &\textbf{0.0305} &0.0543\\
    Gap-I&0.0666 &0.1045* &-- &0.0570 &0.0740 &\textbf{0.0502}\\
    Gap-A&0.0096 &0.0475 &0.0570* &-- &0.0170 &\textbf{0.0068}\\
  \bottomrule
\end{tabular}
\vspace{-0.15in}
\end{table}

Table~\ref{tab:text-evaluation} presents the overall performance. We average the results obtained via FED across its 18 aspects (e.g., interesting, correct, coherent, etc.) on both turn and dialogue levels as the final reported `Score'; `Average' means the average score of the three benchmark datasets; 
`Gap-R' indicate the absolute performance gap between the \underline{R}eDial and each of the rest four, where the smallest gap (best performance) is highlighted in bold and the largest gap (worst performance) is marked with `*'; and `Gap-O, Gap-I, Gap-A' share the similar meanings. From the table, we can notice that no matter which human dialogue dataset is taken as the benchmark, our constructed datasets can achieve a better performance, that is, at least one of ours ranks first (e.g., Coat on Gap-R and Gap-O; ML-1M on Gap-I and Gap-A).
Besides, Figure~\ref{fig:conversation-example} compares text-based conversations in different scenarios, where (a) shows the simulated dialogue in SOTA method HOOPS~\cite{fu2021hoops}; (b) displays our generated dialogue on Coat; and (c) is the real human dialogue on OpenDialKG. 
All of these together help firmly validate the high quality of our generated text-based conversation.

\smallskip\noindent\textbf{Evaluation on the Voice-based Conversation}.
To evaluate the generated voice-based conversation, 
we conduct a user study to compare the voice data generated by the end-to-end VITS with three SOTA two-phase-based TTS models (i.e., Text2Mel + Vocoders): Tacotron 2~\cite{shen2018natural}+WaveGAN~\cite{donahue2019wavegan} (TacoW), Transformer~\cite{li2019neural}+WaveGAN (TranW), FastSpeech 2~\cite{ren2020fastspeech}+WaveGAN (FastW).  
To be specific, we first randomly sample 10 text clips from our generated text-based conversation on Coat and ML-1M, respectively.
Then, we exploit VITS and the three baseline TTS models to generate the corresponding audio clips. Finally, we get 20 groups of audio clips, each associated with one same text clip and four audio clips generated by the four TTS models. 

We evaluate the quality of the four synthesized audio clips within each group based on user studies from three different perspectives:
(1) \textit{obtaining mean opinion score (MOS)~\cite{streijl2016mean} on the naturalness of the audio clips} on a 1-5 scale (bad, poor, fair, good, and excellent); 
(2) \textit{identifying the age of the speaker} with three choices ($<20, [20, 30], >30$); and (3) \textit{identifying the gender of the speaker} with three choices (male, female and others). 
Subsequently, we recruit 10 participants to provide responses to the 20 groups by answering every question. 
To make a fair evaluation without introducing personal bias, we randomly shuffle the four audio clips generated by the four TTS models in each group to avoid some participants consistently rating high
or low for certain audio clips.

\begin{table}[t]
\footnotesize
  \caption{Evaluation on the voice-based conversation. 
  }
  \label{tab:voice-evaluation}
  \addtolength{\tabcolsep}{-0.5pt}
  \vspace{-0.15in}
  \begin{tabular}{l|ccc|ccc}
  \toprule
  &\multicolumn{3}{c|}{Coat} &\multicolumn{3}{c}{ML-1M} \\\cline{2-7}
  &Naturalness &Age &Gender &Naturalness &Age &Gender \\\midrule
  TacoW &3.35 &55.10\% &\textbf{98.98\%} &2.92 &32.00\%* &95.00\%\\
  TranW &3.81 &47.96\%* &97.96\%* &3.60 &34.00\% &95.00\%\\
  FastW &2.87* &55.10\% &98.97\% &2.17* &\textbf{37.00\%} &95.00\%\\
  VITS &\textbf{4.73} &\textbf{58.16\%} &\textbf{98.98\%} &\textbf{4.62} &33.00\% &95.00\%\\
  \bottomrule
\end{tabular}
\vspace{-0.15in}
\end{table}

The MOS on the naturalness of the audio clips and the accuracy of identifying user age $\&$ gender are summarized in Table~\ref{tab:voice-evaluation}, where the best results are highlighted in bold and the worst results are marked by `*'. 
Several interesting observations are noted as follows. 1) The MOS of naturalness on VITS achieves 4.68 on average, which indicates that the synthesized audio is well-qualified according to the criterion. 2) The gender can be easily detected by humans as indicated by the accuracy, viz, the minimal accuracy achieves 95$\%$. 3) In contrast, age is much more difficult to detect by human beings, viz, the accuracy near 50$\%$ on Coat, and around $30\%$ on ML-1M. In summary, these results suggest that extracting side information (such as age) from the generated conversation audio is not trivial for human beings. This inspires us to explore the possible solution using neural networks. 

\begin{figure*}
    \centering \includegraphics[width=0.95\textwidth]{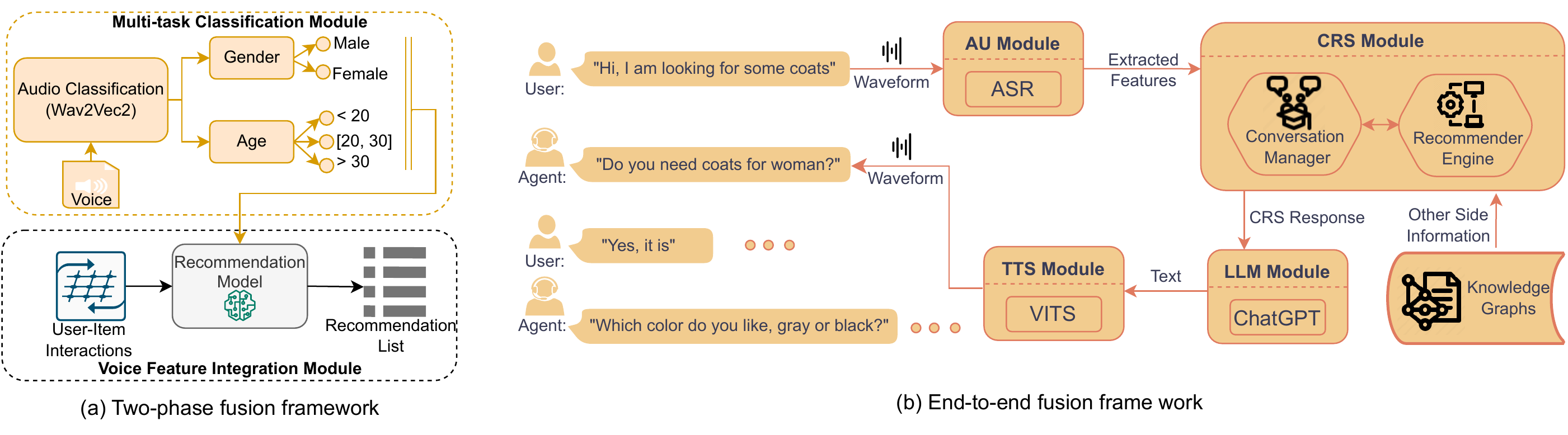}
    \vspace{-0.1in}
    \caption{The overall framework of VCRSs: (a) the two-phase fusion framework; and (b) the end-to-end fusion framework.}
    \label{fig:framework}
    \vspace{-0.15in}
\end{figure*}

\section{Potential Solutions and Directions}
Given our well-constructed VCRS benchmark datasets, this section seeks to (1) point out the challenges of exploiting voice-based data in the context of VCRSs; (2) explore preliminary solutions by using voice-related side information to improve the recommendation performance; and (3) further outline possible directions for building elegant end-to-end VCRSs, with the goal of inspiring more innovative research in this area. 

\begin{table}[t]
\footnotesize
  \caption{Evaluation on user age and gender classification. 
  }
  \label{tab:classification-evaluation}
  \addtolength{\tabcolsep}{2pt}
  \vspace{-0.15in}
  \begin{tabular}{l|cc|cc}
  \toprule
  &\multicolumn{2}{c|}{Coat} &\multicolumn{2}{c}{ML-1M} \\\cline{2-5}
   &Age &Gender &Age &Gender \\\midrule
  The model trained on Coat  &96.70\% &96.98\% &66.50\% &99.40\%\\
  The model trained on ML-1M &81.71\% &83.94\% &84.50\% &91.40\%\\
  \bottomrule
\end{tabular}
\vspace{-0.15in}
\end{table}

\subsection{Challenges for VCRSs}\label{subsec:challenge}
The first major challenge confronted with delivering VCRSs is \textit{how to efficiently extract voice-related features}. As emphasized, voice inputs are capable of conveying richer information than text inputs. For instance, we could possibly identify extra features, such as the user's age, gender, accent, mood, and even context information beyond the conversation content itself. These pieces of information together have been verified to be beneficial for boosting the recommendation performance by both our study (Section~\ref{subsec:advantages-and-necessity}) and previous studies~\cite{covington2016deep,yan2020learning,li2018towards,mairesse2021learning}. Regardless of the advantages, we may take into account several aspects: (1) what features could be extracted from the voice data for better recommendations, for instance, user mood, though being useful, might be hard to distill; and (2) the format of the extracted features, for instance, we might extract either explicit semantic features, e.g., user age and gender from the voice data or implicit latent features represented by low-dimensional embeddings. Moreover, as illustrated in Table \ref{tab:voice-evaluation}, extracting the side information of age is even more difficult for human beings. These are all essential aspects that require a thorough exploration in the context of VCRSs. 

Secondly, \textit{how to seamlessly incorporate the extracted features for performance-enhanced VCRSs} is of great significance. Overall, there could be two possible strategies (1) two-phase fusion and (2) end-to-end fusion. This may be affected by the manner in which these voice-based features are extracted. Specifically, we may first extract explicit semantic features (e.g., user age and gender) from the voice data with the help of separate supervised (e.g., prediction or classification) models, and then fuse these extracted features into the CRS module. This is the so-called two-phase fusion. By contrast, if implicit latent features represented by low-dimensional embeddings are extracted from the voice data, these embeddings can be directly integrated into the CRS module, for example, as part of the state of the reinforcement learning (RL) policy~\cite{zhang2022conversation}, and simultaneously learned with the CRS. This is termed end-to-end fusion. 
However, it remains an open question as to whether the two-phase fusion or the end-to-end fusion is more efficient.
 
\subsection{Preliminary Solution Exploration}
\label{sec:pre-solution}
Bearing the two challenges in mind, we now proceed to explore preliminary solutions for VCRSs. As a starting point, we propose to extract explicit semantic features from the voice data and then incorporate them into the recommendation model in a two-phase fusion manner. Accordingly, our solution consists of two modules, namely the multi-task classification module (MCM) and the voice feature integration module (VFIM) as shown in Figure~\ref{fig:framework}(a).

\smallskip\noindent\textbf{Multi-task Classification Module (MCM)}. 
MCM aims to extract explicit semantic features (e.g., user age) from our created VCRS benchmark datasets. Restricted by the limited information of user/speaker profile, age and gender are the side information that can be extracted from our created VCRS benchmark datasets. To achieve this goal, we first feed our voice-based conversation into the pre-trained ASR model XLSR-Wav2Vec2~\cite{conneau2020unsupervised} to map the voice data into low-dimensional representations, which are subsequently linked with two fully-connected layers to simultaneously perform two tasks, i.e., age prediction and gender prediction.

We train MCM with our two VCRS benchmark datasets, where each dataset is split into training, validation, and test sets with a ratio of 80\%, 10\%, and 10\%. For each dataset, we take the training set to train the MCM, tune the hyper-parameters based on the model performance on the validation set, and finally evaluate the model on the test set. Considering that for each dataset we also use it to train the recommendation model by fusing the prediction via MCM in VFIM, there may be a data leakage issue. To mitigate this, we perform cross-training for MCM, i.e., the MCM trained via one dataset is used to generate predictions for the other in VFIM.

Table~\ref{tab:classification-evaluation} illustrates the prediction accuracy, where we observe that (1) with the non-cross-training setting, MCM achieves better performance on Coat than ML-1M; alternatively stated, age and gender are easier to be identified in the domain of e-commerce (clothing) than movies, which indicates the content of the conversation itself encoded by XLSR-Wav2Vec2 facilitates the inference of age and gender, as female (male) are more inclined to purchase women (men) clothing; (2) MCM performs worse on both datasets with the cross-training setting, which further confirms the fact that the ASR model encodes the content of the conversation, which facilitates the inference of age and gender
to different extents due to domain differences; (3) gender is easier to be correctly predicted than age, which is intuitive as the former is a binary classification task whilst the latter is a multi-class classification one in our study; and (4) comparison on age detection accuracy across Table~\ref{tab:classification-evaluation} and Table~\ref{tab:voice-evaluation} shows that the age classification model can recognize the user age from audio significantly better than human beings, which suggests the side information extraction is a non-trivial task.

\smallskip\noindent\textbf{Voice Feature Integration Module (VFIM)}.
VFIM seeks to integrate the extracted voice-related features into the recommendation model 
for performance enhancements,
which builds a two-phase fusion framework together with the MCM. 
To further examine the efficacy of the 
fused voice features, we adopt the SOTA feature-based recommendation model FM instead of CRS-based 
methods in VFIM. As such, we can compare the two variants: (a) `Age+Gender' - FM trained with age and gender contained in the original Coat and ML-1M datasets shown in Table~\ref{tab:ablation-study} and (b) `Voice Age+Gender' - FM trained with age and gender extracted via MCM from our VCRS benchmark datasets shown in Table~\ref{tab:voice-ablation-study}. With the performance gap between the two variants, we can easily identify the performance degradation caused by the error of MCM. Note that FM can be easily replaced by any existing CRS-based models, e.g., SCPR~\cite{lei2020interactive} and UNICORN~\cite{deng2021unified}. We will keep it as 
our future exploration.

Meanwhile, three additional variants are compared in Table~\ref{tab:voice-ablation-study}: (c) `Voice Gender' and (d) `Voice Age' denoting FM trained with gender and age extracted via MCM from our VCRS benchmark datasets, respectively; and (e) `w/o Age+Gender' denoting FM trained without age and gender. The row `drop-1' indicates the relative performance drop comparing (b) and (a); and `drop-2' shows the relative performance drop comparing (e) and (b). Several findings are noted. \textit{Firstly}, comparing (b) with (a), the average performance drop (3.7702\%, 5.031\% and 6.2206\% w.r.t. Precision, Recall, and NDCG) indicates there is still room to improve MCM thus better easing challenge 1 mentioned in Section~\ref{subsec:challenge}. \textit{Secondly}, comparing (e) with (b), the average performance drop (9.5401\%, 8.4951\% and 6.1150\% w.r.t. Precision, Recall, and NDCG) exhibits the benefits of fusing the extracted voice-related features via MCM, regardless of its imperfect prediction. \textit{Lastly}, both (c) and (d) are defeated by (b), which confirms that both age and gender extracted via MCM from our VCRS benchmark datasets are useful for more accurate recommendations.


\subsection{Future Directions}

The emerging area of VCRS presents opportunities for future research, particularly in the end-to-end pipeline illustrated in Figure~\ref{fig:framework} (b). While we demonstrated the potential of VCRSs in a static context in Section \ref{sec:pre-solution}, a more comprehensive solution would be both end-to-end and dynamic. One promising approach for VCRSs is to integrate a dynamic CRS module into the voice-to-voice pipeline. Specifically, the entire pipeline should include an audio understanding (AU) module that efficiently extracts features from 1) the context information, i.e., text from an automatic speech recognition (ASR) model, and 2) the side information, such as gender, accent, age, and background status. These two types of information can be either disentangled or fused together and fed into a CRS module that can dynamically determine the response for continuing the inquiry or outputting a recommendation list. Moreover, this CRS module could incorporate other side information from external modules, such as knowledge graphs. After generating the response, we can use large language models (LLM) to generate corresponding text sentences. Then, we can use existing text-to-speech (TTS) models to synthesize the response audio and interact with the user repeatedly.

\begin{table}[t]
\centering
\footnotesize
\caption{Performance of FM on Coat and ML-1M, where each experiment is repeatedly executed five times, and the mean $\pm$ standard deviation values (\%) are reported.}
\vspace{-0.15in}
\addtolength{\tabcolsep}{0pt}
\begin{tabular}{ll|ccc}
\toprule
\multicolumn{2}{c|}{} &Precision@10 &Recall@10 &NDCG@10\\\midrule
\multirow{7}{*}{\rotatebox[origin=c]{90}{Coat}} 
&Age+Gender &4.5505 $\pm$ 0.4749 &9.4845  $\pm$ 0.9145 &18.7961 $\pm$ 1.1137\\
&Voice Age+Gender &4.3902 $\pm$ 0.3528 &9.0113 $\pm$ 0.7364 &17.6718 $\pm$ 1.6141\\
&Voice Gender &4.1742 $\pm$ 0.1834 &8.7712 $\pm$ 0.4836 &17.2960 $\pm$ 0.6076\\
&Voice Age &4.3100 $\pm$ 0.3932 &8.9269 $\pm$ 0.8075 &17.5358 $\pm$ 1.6549\\
&w/o Age+Gender &4.1202 $\pm$ 0.3945 &8.6312 $\pm$ 0.8255 &17.2609 $\pm$ 1.5274  \\
&\textit{drop-1} &3.5229\% &4.9892\% &5.9816\%\\
&\textit{drop-2} &6.1501\% &4.2108\% &2.3252\%\\\hline
\multirow{7}{*}{\rotatebox[origin=c]{90}{ML-1M}} 
&Age+Gender  &0.2514 $\pm$ 0.0406 &1.9611 $\pm$ 0.3807 &1.1905 $\pm$ 0.1634\\
&Voice Age+Gender &0.2413 $\pm$ 0.0372 &1.8616 $\pm$ 0.3289 &1.1136 $\pm$ 0.2124\\
&Voice Gender &0.2324 $\pm$ 0.0192 &1.8087 $\pm$ 0.1692 &1.0751 $\pm$ 0.0829\\
&Voice Age &0.2104 $\pm$ 0.0223 &1.6481 $\pm$ 0.2054 &1.0127 $\pm$ 0.1103\\
&w/o Age+Gender &0.2101 $\pm$ 0.0328 &1.6237 $\pm$ 0.3099 &1.0033 $\pm$ 0.1900\\
&\textit{drop-1} &4.0175\% &5.0737\% &6.4595\%\\
&\textit{drop-2} &12.9300\% &12.7793\% &9.9048\%\\
\bottomrule
\end{tabular}
\label{tab:voice-ablation-study}
\vspace{-0.15in}
\end{table}

\section{Conclusion}
This paper focuses on VCRSs because of its user-friendly interaction pattern compared to TCRSs. As there is no existing dataset on VCRSs, the first contribution of this paper is to fill this gap by generating two high-quality VCRS benchmark datasets that future work in this domain can build upon. To do this, we select Coat and ML-1M with rich user \& item features as backbone datasets to simulate text-based conversations, in which ChatGPT generates conversation templates and LightGBM determines the order of asked item features during conversations. On this basis, we use VITS, a neural speech synthesis model, to further generate audio, whereby we utilize user features, such as gender and age, to select a closely matched speaker. We then comprehensively evaluate the quality of the generated VCRS datasets from two perspectives: 1) the quality of text-based conversation is evaluated based on DialoGPT and compared with three real-world conversation datasets; 2) voice-based conversations are evaluated through user studies. 

Additionally, given the generated VCRS datasets, we provide a multi-task classification module to extract the side information, such as age and gender, hidden in the voice conversations. We found that the extracted side information can significantly improve the performance of SOTA recommendation algorithms. Finally, in addition to highlighting the potential of VCRSs, we outline the promising avenues for their future development, such as the dynamic and end-to-end pipeline listed in Figure \ref{fig:framework}. By exploring these directions, we hope to pave the way for further innovation and growth in the field of VCRSs, ultimately enabling these systems to better serve the needs of individuals and organizations.



\begin{acks}
This work is supported by A*STAR Center for Frontier Artificial Intelligence Research and in part by the Data Science and Artificial Intelligence Research Centre, School of Computer Science and Engineering at the Nanyang Technological University, Singapore. 
\end{acks}

\bibliographystyle{ACM-Reference-Format}
\bibliography{reference.bib}


\end{document}